\documentstyle[12pt]{article}
\begin{document}
\def\beq{\begin{equation}}
\def\eeq{\end{equation}}
\def\bea{\begin{eqnarray}}
\def\eea{\end{eqnarray}}
\def\ve{\vert}
\def\vel{\left|}
\def\ver{\right|}
\def\nnb{\nonumber}
\def\ga{\left(}
\def\dr{\right)}
\def\aga{\left\{}
\def\adr{\right\}}
\def\rar{\rightarrow}
\def\nnb{\nonumber}
\def\la{\langle}
\def\ra{\rangle}
\def\ba{\begin{array}}
\def\ea{\end{array}}
\def\tep{$B \rar K \ell^+ \ell^-$}
\def\tepm{$B \rar K \mu^+ \mu^-$}
\def\tept{$B \rar K \tau^+ \tau^-$}
\def\ds{\displaystyle}



\def\bos{\lower 0.5cm\hbox{{\vrule width 0pt height 1.2cm}}}
\def\boss{\lower 0.35cm\hbox{{\vrule width 0pt height 1.cm}}}
\def\aaa{\lower 0.cm\hbox{{\vrule width 0pt height .7cm}}}
\def\dol{\lower 0.4cm\hbox{{\vrule width 0pt height .5cm}}}


\title{ {\Large {\bf 
Exclusive $B \rar \pi \ell^+ \ell^-$ and $B \rar \rho \, \ell^+ \ell^-$
decays in two Higgs doublet model} } }

\author{\vspace{1cm}\\
{\small T. M. Aliev \thanks
{e-mail: taliev@rorqual.cc.metu.edu.tr}\,\,,
M. Savc{\i} \thanks
{e-mail: savci@rorqual.cc.metu.edu.tr}} \\
{\small Physics Department, Middle East Technical University} \\
{\small 06531 Ankara, Turkey} }
\date{}

\begin{titlepage}
\maketitle
\thispagestyle{empty}

\begin{abstract}
\baselineskip  0.7cm
We investigate the exclusive  $B \rar \pi \ell^+ \ell^-$ and $B \rar \rho \,
\ell^+ \ell^-$ decays in framework of the general two Higgs doublet model
(model III), in which an extra phase angle in the charged--Higgs fermion 
coupling, i.e., a new source for CP violation exists. The CP violation for 
both decays are calculated and it is observed that the CP violating asymmetry
in model III 
differs significantly than the one predicted by the standard model and 
model II which is a special case of model III. Furthermore, it is shown that 
the zero value of forward backward asymmetry $A_{FB}$ is shifted when 
compared with the SM value, which can also serve as the efficiency tool 
for establishing new physics.

\end{abstract}

\vspace{1cm}
~~~PACS numbers: 12.60.--i, 13.20.--v, 13.25.Gv
\end{titlepage}

\section{Introduction}
Rare $B$ meson decays, induced by flavor--changing neutral current (FCNC) 
$b \rar s(d)$ transitions, is one of the most promising research area in
particle physics. Theoretical interest to the $B$ meson decays lies in their
role as a potential precision testing ground for the standard model (SM) at
loop level. Experimentally, these decays will provide quantitative
information about the Cabibbo--Kobayashi--Maskawa (CKM) elements
$V_{td},~V_{ts}$ and $V_{tb}$. Besides these rare decay have the potential
for establishing new physics beyond SM, such as two Higgs doublet model
(2HDM), minimal supersymmetric extension of the SM (MSSM) \cite{R1}.

Firstly the most reliable quantitative test of FCNC processes in $B$ decays    
is expected to be measured in inclusive channels. In particular, the decays
$B \rar X_{s,d} \ell^+ \ell^-$ are important probes of the effective
Hamiltonian governing the FCNC transition $b \rar s(d) \ell^+ \ell^-$. The
hope that $B \rar X_s \ell^+ \ell^-$ decay will be measurable in experiments
in the near future, encourage extensive investigation of this process in
the SM, 2HDM and MSSM \cite{R2}--\cite{R15}. The matrix element of the 
$b \rar s \ell^+ \ell^-$ contains terms describing the virtual effects
induced by $t\bar t$,  $c\bar c$  and $u\bar u$ loops which are proportional
to $V_{tb}\,V_{ts}^*$, $V_{bc}\,V_{cs}^*$ and $V_{bu}\,V_{su}^*$,
respectively. Using unitarity of the CKM matrix and neglecting
$V_{bu}\,V_{su}^*$ in comparison to $V_{tb}\,V_{ts}^*$ and
$V_{bc}\,V_{cs}^*$, it is obvious that the matrix element for the 
$b \rar s \ell^+ \ell^-$ involves only one independent CKM factor 
$V_{tb}\,V_{ts}^*$ so that CP--violation in this channel is strongly
suppressed in the SM. 

The situation is totally different for the $b \rar d \ell^+ \ell^-$ decay,
since all three CKM factors $V_{tb}\,V_{td}^*$, $V_{cb}\,V_{cd}^*$ and
$V_{ub}\,V_{ud}^*$, are all of the same order (in SM) and therefore can
induce considerable CP--violating difference  between the decay rates of the 
reactions $b \rar d \ell^+ \ell^-$ and $\bar b \rar \bar d \ell^+ \ell^-$.

It should be noted here that in presence of a much stronger decay 
$b \rar s \ell^+ \ell^-$, the detection of the $b \rar d \ell^+ \ell^-$
decay seems to be more problematic. For this reason, in search of
CP violation the corresponding exclusive decay channels 
$B \rar \pi \ell^+ \ell^-$ and $B \rar \rho\, \ell^+ \ell^-$ are more
preferable.
In general, the inclusive decays are rather difficult to measure in
comparison to the exclusive ones. CP--violating effects in inclusive
$b \rar d \ell^+ \ell^-$ and exclusive $B \rar \pi \ell^+ \ell^-$, 
$B \rar \rho\, \ell^+ \ell^-$ channels were studied within the framework of
the SM in \cite{R15,R16}.

The aim of the present work is to derive quantitative predictions for the 
CP violation in the exclusive $B \rar \pi \ell^+ \ell^-$ and 
$B \rar \rho\, \ell^+ \ell^-$ decays, in context of the general two Higgs doublet
model, in which a new source for CP violation is present (see below). 2HDM
model is one of the simplest extension of the SM, which contains 
two complex Higgs doublets, while the SM contains only one. In general, 
in 2HDM the flavor changing neutral currents (FCNC) that appear at tree 
level, are avoided by imposing an {\it ad hoc} discrete symmetry \cite{R18}. 
One possible approach to avoid these unwanted FCNC at tree level is to
couple all fermions to only one of the above--mentioned Higgs doublets
(model I). The other possibility is the coupling of the up and down quarks
to the first and second Higgs doublets, with the vacuum expectation values 
$v_2$ and $v_1$, respectively (model II). Model II is more attractive since
its Higgs sector coincides with the ones in the supersymmetric model.
In this model there exist five physical Higgs fields: neutral
scalars $H^0$, $h^0$, neutral pseudoscalar $A$ and charged Higgs bosons
$H^\pm$. The interaction vertex of fermions with Higgs fields depends on
$tan \beta=v_2/v_1$, which is the free parameter of the model. The new
experimental results of CLEO and ALEPH Collaborations \cite{R19,R20} 
on the branching ratio
$b \rar s \gamma$ decay impose strict restrictions on the charged Higgs
boson mass and $tan \beta$. Recently, the lower bound on these parameters
were determined from the analysis of the $b \rar s \gamma$ decay, including            
NLO QCD corrections \cite{R21,R22}. Other indirect bound on the ratio
$m_{H^\pm}/tan \beta$ come from $B \rar D \tau \bar \nu_\tau$ decay, where
$m_{H^\pm} \ge 2.2 tan \beta~GeV$ \cite{R23}, and from
the $\tau$ lepton decays $m_{H^\pm} \ge 1.5 \, tan \beta ~GeV$ 
\cite{R24}. The consequence of an
analysis without discrete symmetry has been investigated in a more general
model in 2HDM, namely, model III \cite{R25,R26}. In this model FCNC appears 
naturally at tree level. However, the FCNC's involving the first two
generations are highly suppressed, as is observed in the low energy
experiments, and those involving the the third generation is not as severely
suppressed as the first two generations, which are restricted by the
existing experimental results. 

In this work we assume that all tree level FCNC couplings are negligible.
It should be noted however that, even with this assumption, the couplings of 
fermions to Higgs bosons may have a complex phase $e^{i\theta}$. 
In other words, in this model there exists a new source of CP violation that
is absent in the SM, model I and model II.
The effects of such an extra phase in the $b \rar s \gamma$ decay were discussed
in \cite{R27,R28}. The constraints on the phase angle $\theta$ in the product 
$\lambda_{tt} \lambda_{bb}$ of Higgs--fermion coupling (see below) 
imposed by the neutron electric dipole moment, 
$B^0 - \bar B^0$ mixing. $\rho \,_0$ parameter and $R_b$ is discussed in 
\cite{R28}.        

The paper is
organized as follows: In Section 2 we present the necessary theoretical
framework. The branching ratios, CP--violating effects in the partial
widths and forward--backward asymmetry for the above--mentioned 
exclusive decay channels are studied in section 3. Section 4 is devoted to
the numerical analysis and concluding remarks.       
  
\section{Theoretical framework}

Before presenting the necessary theoretical background, let us go through
the main essential points of the general Higgs doublet model (model III). 
In this model, both Higgs
doublets can couple to up and down quarks. Without loss of generality we can
work in a basis such that the first doublet generates all the fermion and
gauge boson masses, whose vacuum expectation values are
\bea
\left< \phi_1 \right> = \left( \begin{array}{c}
0 \\ \\
\displaystyle{\frac{v}{\sqrt{2}}}	
\end{array} \right)~~~~~,~ \left< \phi_2 \right>~. \nnb
\eea
In this basis the first doublet $\phi_1$ is the same as in the SM, and all
new Higgs bosons result from the second doublet $\phi_2$, which can be
written in the following form
\bea
\phi_1 = \frac{1}{\sqrt{2}} \left( \begin{array}{c}
\sqrt{2}\, G^+ \\ \\
v + \chi_1^0 + i G^0 
\end{array} \right)~~~~~,
~ \phi_2  = \frac{1}{\sqrt{2}} \left( \begin{array}{c}
\sqrt{2}\, H^+ \\ \\
\chi_2^0 + i A^0 
\end{array} \right)~, \nnb
\eea
where $G^+$ and $G^0$ are the Goldstone bosons. The neutral $\chi_1^0$ and
$\chi_2^0$ are not the physical mass eigenstate, but their linear
combinations give the neutral $H^0$ and $h^0$ Higgs bosons:
\bea
\chi_1^0 = H^0 cos \alpha - h^0 sin \alpha~, \nnb \\
\chi_2^0 = H^0 sin \alpha + h^0 cos \alpha~. \nnb
\eea
The general Yukawa Lagrangian can be written as 
\bea
{\cal L}_Y = \eta_{ij}^U \bar Q_{iL} \tilde \phi_1 U_{jR} +
\eta_{ij}^{\cal D} \bar Q_{iL} \phi_1 {\cal D}_{jR}
+ \xi_{ij}^U \bar Q_{iL} \tilde \phi_2  U_{jR} +
\xi_{ij}^{\cal D} \bar Q_{iL} \phi_2  {\cal D}_{jR} + h.c.~,
\eea
where $i$, $j$ are the generation indices, $\tilde \phi= i \sigma_2 \phi$,
$\eta_{ij}^{U,{\cal D}}$ and $\xi_{ij}^{\cal U,D}$, in general, are the
non--diagonal coupling matrices, $L=(1-\gamma_5)/2$ and $R=(1+\gamma_5)/2$ are
the left-- and right--handed projection operators. In Eq. (1) all states
are weak states, that can be transformed to the mass eigenstates by
rotation. After performing this rotation on the Yukawa Lagrangian, we get
\bea
{\cal L}_Y = - H^+ \bar U \left[ V_{CKM} \hat \xi^{\cal D} R - 
\hat \xi^{U^+} V_{CKM} L \right] {\cal D}~,
\eea
where $U({\cal D})$ represents the mass eigenstates of $u,~c,~t~(d,~s,~b)$
quarks. In the present analysis, we will use a simple ansatz for
$\hat \xi^{U^+,{\cal D}}$ \cite{R25},
\bea
\hat \xi^{U^+,{\cal D}} = \lambda_{ij} 
\frac{g \sqrt{m_i m_j}}{\sqrt{2} m_W}~.
\eea
Also it assumed that $\lambda_{ij}$ is complex, i.e., 
$\lambda_{ij} = \vel \lambda_{ij} \ver e^{i\theta}$, and for simplicity we
choose $\xi^{U,{\cal D}}$ to be diagonal to suppress all tree level FCNC
couplings, and as a result $\lambda_{ij}$ are also diagonal but remain
complex. Note that the results for model I and model II can be obtained from
model III by the following substitutions:
\bea 
&&\lambda_{tt} = cot \beta ~~~~~~~ \lambda_{bb} = - \, cot \beta ~~\mbox{\rm
for model I}~, \nnb \\
&&\lambda_{tt} = cot \beta ~~~~~~~ \lambda_{bb} = + \, tan  \beta ~~\mbox{\rm   
for model II}~,
\eea
and $\theta = 0$.

After this brief introduction about the general Higgs doublet model,
let us return our attention to the $b \rar d \ell^+ \ell^-$
decay. The powerful framework into which the 
perturbative QCD corrections to the physical decay amplitude incorporated 
in a systematic way, is the effective Hamiltonian method. 
In this approach, the heavy degrees of freedom in the present case, 
i.e., $t$ quark, $W^\pm,~H^\pm,~h^0,~H^0$ are integrated out. 
The procedure is to match the
full theory with the effective theory at high scale $\mu = m_W$, and then
calculate the Wilson coefficients at lower $\mu \sim {\cal O}(m_b)$ using the 
renormalization group equations. In our calculations we choose the higher
scale as $\mu = m_W$, since the charged Higgs boson is heavy enough
($m_{H^\pm} \ge 210 ~GeV$ see \cite{R21}) to neglect the evolution from 
$m_{H^\pm}$ to $m_W$. 

In the version of the 2HDM we consider in this work, the charged Higgs boson 
exchange diagrams do not
produce new operators and the operator basis is the same as the one used 
for the $b \rar d \ell^+ \ell^-$ decay in the SM.
For this reason in the model under consideration, 
the charged Higgs boson contributions to leading order change only the value
of the Wilson coefficients at $m_W$ scale, i.e.,
\bea
C_7^{2HDM}(m_W) &=& C_7^{SM}(m_W) + C_7^{H^\pm}(m_W) \nnb \\
C_9^{2HDM}(m_W) &=& C_9^{SM}(m_W) + C_9^{H^\pm}(m_W) \nnb \\
C_{10}^{2HDM}(m_W) &=& C_{10}^{SM}(m_W) + C_{10}^{H^\pm}(m_W)~. \nnb
\eea
The coefficients $C_i^{2HDM} (m_W)$ to the leading order are given by
\bea
C_7^{2HDM}(m_W) &=& 
x \, \frac{(7-5 x - 8 x^2)}{24 (x-1)^3} + 
\frac{x^2 (3 x - 2)}{4 (x-1)^4} \, \ell n x \nnb \\
&+& \vel \lambda_{tt} \ver^2 \Bigg( \frac{y(7-5 y - 8 y^2)}
{72 (y-1)^3} + \frac{y^2 ( 3 y - 2)}{12 (y-1)^4} \, \ell n y \Bigg) \nnb \\
&+& \lambda_{tt} \lambda_{bb} \Bigg( \frac{y(3-5 y)}{12 (y-1)^2} +
\frac{y (3 y - 2)}{6 (y-1)^3} \, \ell n y \Bigg)~, \\ \nnb \\ \nnb \\
C_9^{2HDM}(m_W) &=& - \frac{1}{sin^2 \theta_W} \, B(m_W) + 
\frac{1 - 4 sin^2 \theta_W}{sin^2 \theta_W} \, C(m_W) \nnb \\
&+& \frac{-19 x^3 + 25 x^2}{36 (x-1)^3} +
\frac{-3 x^4 + 30 x^3 - 54 x^2 + 32 x -8}{18 (x-1)^4} \, \ell n x 
+ \frac{4}{9} \nnb \\
&+& \vel \lambda_{tt} \ver^2 \Bigg[ 
\frac{1 - 4 sin^2 \theta_W}{sin^2 \theta_W} \, \frac{x y}{8} \Bigg( 
\frac{1}{y-1} - \frac{1}{(y-1)^2} \, \ell n y \Bigg)\nnb \\
&-& y \Bigg( \frac{47 y^2 - 79 y + 38}{108 (y-1)^3}
-\frac{3 y^3 - 6 y^3 + 4}{18 (y-1)^4} \, \ell n y \Bigg) \Bigg]~,
\\ \nnb \\ \nnb \\
C_{10}^{2HDM}(m_W) &=& \frac{1}{sin^2 \theta_W} \Big( B(m_W) - 
C(m_W) \Big) \nnb \\
&+& \vel \lambda_{tt} \ver^2 \frac{1}{sin^2 \theta_W} \,\frac{x y}{8} 
\Bigg( - \frac{1}{y-1} + \frac{1}{(y-1)^2} \, \ell n y \Bigg)~, 
\eea
where
\bea
B(x) &=& - \frac{x}{4 (x-1)} + \frac{x}{4 (x-1)^2} \, \ell n x ~, \nnb \\
C(x) &=& - \frac{x}{4} \Bigg( \frac{x-6}{3 (x-1)} + 
\frac{3 x +2 }{2 (x-1)^2} \ell n x \Bigg)~,\nnb \\
x &=& \frac{m_t^2}{m_W^2} ~, \nnb \\
y &=& \frac{m_{H^\pm}^2}{m_W^2}~.
\eea
and $sin^2\theta_W = 0.23$ is the Weinberg angle. 
It follows from Eqs. (5--7)
that among all the Wilson coefficients, only $C_7$ involves the new phase
angle $\theta$.
We have neglected the
neutral Higgs boson exchange diagram contributions, since Higgs boson--fermion
interaction is proportional to the lepton mass. 

The effective Hamiltonian for the $b \rar d \ell^+ \ell^-$ decay is 
[29--32]
\bea
{\cal H} = -4 \frac{G_F}{2\sqrt 2} V_{tb} V^*_{td}
\left\{\sum_{i=0}^{10} C_i(\mu) O_i(\mu) +
\lambda_u \sum_{i=1}^{2} C_i(\mu) \left[ O_i(\mu) - O_i^u(\mu)
\right] \right\}~, \nnb
\eea
where 
$$ \lambda_u = \ds{\frac{V_{ub} V^*_{ud}}{V_{tb} V^*_{td}}}~,$$ and $C_i$ are
the Wilson coefficients.
The explicit form of all operators $O_i$ can be found in [29--32].

The evolution of the Wilson coefficients from the higher scale 
$\mu = m_W$ down to the low energy scale $\mu = m_b$ is described by the
renormalization group equation
\bea
\mu \frac{d}{d\mu}C_i^{eff(\mu)} = 
C_i^{eff}(\mu) \gamma_\mu^{eff}(\mu)~, \nnb
\eea
where $\gamma$ is the anomalous dimension matrix. 
The coefficient $C_7^{eff}(\mu)$ at the scale ${\cal O}(m_b)$ 
in NLO is calculated in \cite{R21,R22}:
\bea
C_7^{eff}(m_b) = C_7^0(m_b) + \frac{\alpha_s(m_b)}{4 \pi} 
C_7^{1,eff}(m_b)~,\nnb
\eea
where $C_7^0(m_b)$ is the leading order (LO) term and $C_7^{1,eff}(m_b)$
describes the NLO terms, whose explicit forms can be found in 
\cite{R21}. In our case,
the expressions for these coefficients can be obtained from the results of 
\cite{R21} by making the following replacements:
\bea
\vel Y \ver^2  \rar \vel \lambda_{tt} \ver^2 ~~~~~ \mbox{\rm and} ~~~~~ 
X Y^*  \rar \vel \lambda_{tt} \lambda_{bb} \ver e^{i\theta}~. \nnb
\eea
In the SM, the QCD corrected Wilson coefficient $C_9(m_b)$, which
enters to the decay amplitude up to the next leading order has been
calculated in [29--32]. The Wilson coefficient $C_{10}$ does not 
receive any new corrections at all, i.e., $C_{10}(m_b) \equiv 
C_{10}^{2HDM}(m_W)$.  As we have already noted, in the version of the 2HDM we
consider in this work, there does not appear any new operator other than
those that exist in the SM, therefore it is enough to make the
replacement $C_9^{SM}(m_W) \rar C_{9}^{2HDM}(m_W)$ in [29--32], in order
to calculate $C_{9}^{2HDM}$ at $m_b$ scale. Hence, including the NLO
QCD corrections, $C_9(m_b)$ can be written as:
\bea
\lefteqn{
C_9(\mu) = C_9^{2HDM}(\mu) 
\left[1 + \frac{\alpha_s(\mu)}{\pi} \omega (\hat s) \right]} \nnb \\
&&+ \, g(\hat m_c,\hat s) \Big[ 3 C_1(\mu) + C_2(\mu) + 3 C_3(\mu) + C_4(\mu)
+ 3 C_5(\mu) + C_6(\mu) \Big] \nnb \\
&&+ \,\lambda_u \Big[g(\hat m_c,\hat s) - g(0,\hat s)\Big] 
\Big( 3 C_1(\mu) + C_2(\mu) \Big) - 
\frac{1}{2} g(0,\hat s) \ga C_3(\mu) + 3 C_4(\mu) \dr \nnb \\
&&-\,  \frac{1}{2} g \ga 1, \hat s\dr
\ga 4 C_3 + 4 C_4 + 3 C_5 + C_6 \dr
- \frac{1}{2} g \ga 0, \hat s\dr \ga C_3 + 3 C_4 \dr \nnb \\
&&+\, \frac{2}{9} \ga 3 C_3 + C_4 + 3 C_5 + C_6 \dr~, 
\eea
where $m_c = m_c/m_b~, ~\hat s = p^2/m_b^2$, and
\bea
\lefteqn{
\omega \ga \hat s \dr = - \frac{2}{9} \pi^2 - 
\frac{4}{3} Li_2  \ga \hat s \dr - \frac{2}{3} \ell n \ga \hat s\dr 
\,\ell n \ga 1 -\hat s \dr} \nnb \\ 
&&- \,\frac{5 + 4 \hat s}{3 \ga 1 + 2 \hat s \dr} \ell n \ga 1 -\hat s \dr
-\frac{2 \hat s \ga 1 + \hat s \dr \ga 1 - 2 \hat s \dr}
{3 \ga 1 - \hat s \dr^2 \ga 1 + 2 \hat s \dr} \, \ell n \ga \hat s\dr 
+ \frac{5 + 9 \hat s - 6 {\hat s}^2}
{3 \ga 1 - \hat s \dr \ga 1 + 2 \hat s \dr}~
\eea
represents the ${\cal O}\ga \alpha_s \dr$ correction from the one gluon
exchange in the matrix element of $O_9$, while the function
$g \ga \hat m_c, \hat s \dr$ arises from one loop contributions of the
four--quark operators $O_1$--$O_6$, whose form is
\bea 
\lefteqn{
g \ga \hat m_c, \hat s \dr = - \frac{8}{9} \ell n \ga \hat m_i\dr
+ \frac{8}{27} + \frac{4}{9} y_i - \frac{2}{9} \ga 2 + y_i \dr} \nnb \\
&& +\, \sqrt{\vel 1-y_i \ver} \Bigg\{ \Theta \ga 1 - y_i \dr 
\Bigg( \ell n \frac{1+\sqrt{\vel 1-y_i \ver}}{1-\sqrt{\vel 1-y_i \ver}}
- i \, \pi \Bigg)
+ \Theta \ga y_i -1 \dr 2 arctan \frac{1}{\sqrt{y_i - 1}} \Bigg\}~,
\eea
where $y_i = 4 {\hat m_i}^2/{\hat p}^2$.

The Wilson coefficients $C_9$ receives also long distance contributions,
which have their origin in the real $u\bar u$, $d\bar d$ and $c\bar c$ 
intermediate states, i.e., $\rho$, $\omega$ and $J/\psi$, $\psi^\prime$,
$\cdots$. In the case of the $J/\psi$ family this usually
accomplished by introducing a Breit--Wigner distribution for the resonance
through the replacement ([4--7,33])
\bea
g \ga \hat m_c, \hat s \dr \rar g \ga \hat m_c, \hat s \dr -
\frac{3\pi}{\alpha^2_{em}}
\, \kappa \sum_{V_i=J/\psi_i,\psi^\prime,\cdots} 
\frac{m_{V_i} \Gamma(V_i \rar \ell^+ \ell^-)}
{(p^2 - m_{V_i}^2) + i m_{V_i} \Gamma_{V_i}}~,
\eea
where the phenomenological parameter $\kappa =2.3$ is chosen in order to
reproduce correctly the experimental value of the branching ratio
(see for example \cite{R16})
\bea
{\cal B}(B \rar J/\psi X \rar X \ell^+ \ell^-)={\cal B}(B \rar J/\psi X)   
\,{\cal B}(J/\psi \rar X \ell^+ \ell^-)~. \nnb
\eea
In order to avoid the double counting,
in this work,  as an alternative to the functions 
$g \ga \hat m_u, \hat s \dr$ and $g \ga \hat m_c, \hat s \dr$ that
describe the effects of $u \bar u$ and $c \bar c$ loops, we have used a
different procedure, in which these functions are expressed through the
normalized vacuum polarization $\Pi^\gamma_{had}(\hat s)$ that is related to
the experimentally measurable quantity 
\bea
R_{had}(\hat s) = \frac{\sigma_{tot} (e^+ e^- \rar hadrons)}
{\sigma(e^+ e^- \rar \mu^+ \mu^-)}~, 
\eea  
via the dispersion relation ( see \cite{R16,R17} for more detail).
In this way it is possible to include the
$\rho,~\omega,~J/\psi,~\psi^\prime,~\cdots$ resonances into the differential
cross section in an approximate way, consistent with the idea of global
duality. In this approach the $\omega$ and $J/\psi$ family resonances are
well described through the Breit--Wigner form and $\rho$ resonance is
introduced by
\bea
R_{res}^\rho = \frac{1}{4} \ga 1 - 4 \frac{\hat m_\pi^2}{\hat s} \dr^{3/2}
\vel F_\pi (\hat s) \ver^2~,
\eea
where $F_\pi (\hat s)$ is the pion form factor that is represented by a
modified Gounaris--Sakurai formula (see \cite{R34,R35}).

The effective short--distance Hamiltonian for $b \rar d\ell^+ \ell^-$
decay [29--32] leads to the QCD corrected matrix element (when the $d$
quark mass is neglected)
\bea
{\cal M} &=& \frac{G_F \alpha}{2\sqrt 2 \pi} V_{td} V^*_{tb} \Bigg\{
C_9^{eff} \bar d \gamma_\mu (1- \gamma_5) b \, \bar \ell \gamma^\mu \ell + 
C_{10} \bar d \gamma_\mu (1- \gamma_5) b \, \bar \ell 
\gamma^\mu \gamma_5 \ell \nnb \\
&-& 2 C_7\frac{m_b}{p^2}\bar d i \sigma_{\mu \nu}p^\nu (1+\gamma_5)  b  \,
\bar \ell \gamma^\mu \ell~,
\eea
where $p^2$  is the invariant dilepton mass. In Eq. (12) all Wilson
coefficients are evaluated at the $\mu = m_b$ scale.
 
\section{The exclusive $B \rar \pi \ell^+ \ell^-$ and 
$B \rar \rho\, \ell^+ \ell^-$ decays}  
In this section, we proceed to calculate the
branching ratio and CP violating asymmetry in the 
$B \rar \pi \ell^+ \ell^-$ and $B \rar \rho\, \ell^+ \ell^-$ decays. It
follows from the matrix element of the $b \rar d \ell^+ \ell^-$ that 
in order to be able to calculate the matrix element of the exclusive decay
$B \rar M \ell^+ \ell^-$, the matrix elements 
$\la M \ve \bar d \gamma_\mu (1 + \gamma_5 ) b \ve B \ra$ and 
$\la M \ve \bar d i\, \sigma_{\mu\nu}p_\nu (1 + \gamma_5 ) b \ve B \ra$ 
($M=\pi$ or $\rho$)
have to be calculated. These matrix elements can be parametrized in the
following way:
\bea
\left< \pi \ga p_\pi \dr \vel \bar d \gamma_\mu (1 - \gamma_5 ) b 
\ver B \ga p_B \dr \right> &=& 
f^+(p^2) \ga p_B + p_\pi \dr_\mu + f^-(p^2) p_\mu~, \\
\left< \pi \ga p_\pi \dr \vel \bar d i\, \sigma_{\mu\nu}p_\nu (1 + \gamma_5 ) b 
\ver B \ga p_B \dr \right> &=& 
\left[\ga p_B + p_\pi \dr_\mu p^2 - p_\mu \ga m_B^2 - m_\pi^2 \dr \right]
\frac{f_T(p^2)}{m_B+m_\pi}~, \\
\left< \rho (p_\rho, \varepsilon) \vel \bar d \gamma_\mu( 1- \gamma_5) b \ver
B(p_B) \right> &=&
- \epsilon_{\mu \nu \lambda \sigma} \varepsilon^{*\nu} p_\rho^\lambda p_B^\sigma
\frac{2 V(p^2)}{m_B + m_{\rho}} \nnb \\ 
&-& i \varepsilon_\mu^* ( m_B + m_\rho) A_1(p^2)
+ i (p_B + p_\rho)_\mu (\varepsilon^* p) \frac{A_2(p^2)}{m_B + m_\rho}\nnb \\
&+& i p_\mu (\varepsilon^* p) \frac{2 m_\rho}{p^2} \left[ A_3(p^2) - A_0 (p^2)
\right] ~, \\ 
\left< \rho (p, \varepsilon) \vel \bar d i \sigma_{\mu \nu} p^\nu (1+ \gamma_5) b
\ver B(p_B) \right> &=&
4 \epsilon_{\mu \nu \lambda \sigma} \varepsilon^{* \nu} p^\lambda
p^\sigma T_1 (p^2) \nnb \\
&+&  2 i \left[ \varepsilon_\mu^* (m_B^2 - m_\rho^2) - (p_B + p_\rho)_\mu (\varepsilon^* p)
\right] T_2 (p^2) \nnb \\
&+&  2 i (\varepsilon^* p) \left[ p_\mu - (p_B+p_\rho)_\mu \frac{p^2}{m_B^2 -
m_\rho^2} \right] T_3 (p^2)~. 
\eea
In all of the matrix elements above, 
$p = p_B-p_M$ ($M = \pi$ or $\rho$) and $\varepsilon^*$ is the 
four--polarization vector of the $\rho$ meson.
Using Eqs. (15--19) we obtain for the matrix elements of the 
$B \rar \pi \ell^+ \ell^-$ and $B \rar \rho\, \ell^+ \ell^-$ decays:
\bea
{\cal M}^{B \rar \pi} &=& \frac{G \alpha}{2 \sqrt{2} \pi}\,
V_{tb} V_{td}^* \Bigg\{ \ga 2 A p_{\pi\mu} + B p_\mu \dr
\bar \ell \gamma_\mu \ell 
+\ga 2 C p_{\pi\mu} + D p_\mu \dr\bar \ell \gamma_\mu \gamma_5 \ell
\Bigg\}~, \\ \nnb \\ \nnb \\
{\cal M}^{B \rar \rho} &=& \frac{G \alpha}{2 \sqrt{2} \pi} V_{tb} V_{td}^*  
\Bigg\{ \bar \ell \gamma_\mu \ell 
\left[ 2 A_1 \epsilon_{\mu \nu \lambda \sigma} 
\varepsilon^{* \nu} p_\rho^\lambda p_B^\sigma +
i B_1 \varepsilon^*_\mu - i B_2 ( \varepsilon^* p) ( p_B + p_\rho)_\mu 
- i B_3 (\varepsilon^* p) p_\mu \right] \nnb \\
&+& \bar \ell \gamma^\mu \gamma_5 \ell 
\left[ 2 C_1 \epsilon_{\mu \nu \lambda \sigma}
\epsilon^{* \nu} p_\rho^\lambda p_B^\sigma + 
i D_1 \epsilon^*_\mu - i D_2 (\varepsilon^* p) ( p_B + p_\rho)_\mu - 
i D_3 (\varepsilon^* p) p_\mu \right] \Bigg\}~,
\eea
where
\bea
A &=& C_9^{eff} f^+ - C_7 \frac{2 m_b f_T(p^2)}{m_B + m_\pi}~, \nnb \\
B &=& C_9^{eff} \ga f^+ + f^- \dr + C_7 \ga \frac{2 m_b f_T}{p^2} \dr
\ga \frac{m_B^2 - m_\pi^2 - p^2}{m_B+m_\pi}\dr~,\nnb \\
C &=& C_{10} f^+ ~, \nnb \\
D &=& C_{10} \ga f^+ + f^- \dr ~, \nnb \\
A_1 &=& C_9^{eff} \frac{V}{m_B + m_\rho} + 4 C_7 \frac{m_b}{p^2} T_1~, \nnb \\ 
B_1 &=& C_9^{eff} (m_B + m_\rho) A_1 + 4 C_7 \frac{m_b}{p^2} (m_B^2 -
m_\rho^2) T_2~,  \nnb \\ \nnb \\ 
B_2 &=& C_9^{eff} \frac{A_2}{m_B + m_\rho} + 4 C_7 \frac{m_b}{p^2} \ga T_2
+
 \frac{p^2}{m_B^2 - m_\rho^2} T_3 \dr~,  \nnb \\ \nnb \\
B_3 &=& -C_9^{eff}\frac{ 2 m_\rho}{  p^2}(A_3 - A_0) + 4 C_7
\frac{m_b}{p^2}T_3~,  \\ \nnb \\
C_1 &=& C_{10} \frac{V}{m_B + m_\rho}~,  \nnb \\ \nnb \\
D_1 &=& C_{10} (m_B + m_\rho) A_1~,  \nnb \\ \nnb \\  
D_2 &=& C_{10} \frac{A_2}{m_B + m_\rho}~,  \nnb \\ \nnb \\
D_3 &=& C_{10} \frac{2 m_\rho}{p^2} (A_3 - A_0)~. \nnb
\eea
Using Eqs. (20) and (21) and performing summation over final lepton
and $\rho$ meson polarization (in the $B \rar \rho\, \ell^+ \ell^-$ case),
we obtained the following results for the double differential decay rates
(the masses of the leptons, in our case electron or muon, are neglected):
\bea
\lefteqn{
\frac{d \Gamma^{B \rar \pi}}{dp^2 dz} = \frac{G^2 \alpha^2 }{2^{11} \pi^5} \, 
\frac{ \vel V_{tb} V_{ts}^* \ver^2 \sqrt \lambda}{m_B} 
\,\lambda m_B^4 (1-z^2) \Bigg[ \ga \vel A \ver^2 +
\vel C \ver^2 \dr \Bigg]~,} \\ \nnb \\ \nnb \\ 
\lefteqn{
\frac{d \Gamma^{B \rar \rho}}{d p^2 dz} = \frac{G^2 \alpha^2  
\vel V_{tb} V_{td}^* \ver^2\sqrt{\lambda}}{2^{12}
\pi^5 m_B} \Bigg\{ 2 \lambda m_B^4 \Bigg[ m_B^2 s ( 1+ z^2) \ga 
\vel A_1 \ver ^2 +\vel C_1 \ver ^2 \dr \Bigg] } \nnb \\
&&+ \frac{1}{2 r} \Bigg[ m_B^2 \ga \lambda (1- z^2) + 8 r s \dr
\ga \vel B_1 \ver^2 + \vel D_1 \ver^2 \dr
- 2 \lambda m_B^4
(1-r-s)(1-z^2)  \nnb \\ && \times \Big( Re\ga  B_1 B_2^*  \dr + Re \ga D_1 D_2^*  \dr      
\Big)\Bigg]
+ \lambda^2 m_B^6 (1- z^2) \, \frac{1}{2 r}\, \ga \vel B_2
\ver ^2 + \vel D_2 \ver ^2 \dr \nnb \\
&&+ 8 m_B^4 s z \sqrt{\lambda} \Big( Re\ga  B_1 C_1^* \dr + Re\ga A_1 D_1^*\dr \Big)
 \Bigg\}~,
\eea
where $z=cos \theta$\,, $\theta$ is the angle between the three--momentum of the
$\ell^+$ lepton and that of the $B$ meson in the center of mass frame of the 
lepton pair,
$\lambda(1,r_M,s) = 1+r_M^2+s^2 -2 r_M - 2 s - 2 r_M s$, $r_M =
\frac{\ds{m_M^2}}{\ds{m_B^2}}$, and 
$s=\frac{\ds{p^2}}{\ds{m_B^2}}$ ($M=\pi$ or $\rho$).
The CP violating asymmetry between $B \rar M \ell^+ \ell^-$ 
and $\bar B \rar \bar M \ell^+ \ell^-$ decays is defined as 
\bea
A_{CP} (p^2) = \frac{\displaystyle{\frac{d \Gamma}{dp^2} -
\frac{d \bar \Gamma}{dp^2}}}{\displaystyle{\frac{d \Gamma}{dp^2}+
\frac{d \bar \Gamma}{dp^2}}}~.
\eea
where 
\bea
\frac{d \Gamma}{dp^2} = \frac{d \Gamma \ga \bar B \rar M \ell^+ \ell^- \dr}
{dp^2}~~~~~~~~~~\mbox{\rm and}~~~~~~~~~~
\frac{d \bar \Gamma}{dp^2} = \frac{d \Gamma \ga  B \rar \bar M \ell^+ \ell^-
\dr} {dp^2}~. \nnb
\eea
The differential decay widths $B \rar \pi \ell^+ \ell^-$ and 
$B \rar \rho\, \ell^+ \ell^-$ can easily be obtained from Eqs. (23) and
(24) by integrating over $z$. Finally we get the following results
for CP violating asymmetry for the $B \rar \pi \ell^+ \ell^-$ and
$B \rar \rho\, \ell^+ \ell^-$ decays
\bea
A_{CP}^{B \rar \pi}(p^2) &\simeq& - \frac{2}{\ga \vel A \ver^2
+ \vel C \ver^2 \dr}
\Bigg\{ \vel f_+ \ver^2 \ga \mbox{\rm Im} \lambda_u \dr
\ga \mbox{\rm Im}\xi_1^* \xi_2 \dr \\
&+& f_+ f_T \, \frac{2 m_b}{m_B + m_\pi} 
\Big[ \ga \mbox{\rm Im} \xi_1 \dr \eta_2 -
\ga \mbox{\rm Im} \lambda_u \dr \ga \mbox{\rm Im} \xi_2 \dr \eta_1 
+ \ga \mbox{\rm Re} \lambda_u \dr \ga  \mbox{\rm Im} \xi_2 \dr \eta_2 
\Big]\Bigg\}~, \nnb \\ \nnb \\ \nnb \\
A_{CP}^{B \rar \rho}(p^2) &\simeq& {\ds \frac{1}{\Sigma^\rho}} \, \Bigg\{ - 2 
\ga \mbox{\rm Im} \lambda_u \dr \ga \mbox{\rm Im}\xi_1^* \xi_2 \dr
\Bigg[ \frac{16}{3} \lambda m_B^6 s \vel \frac{V}{m_B+m_\rho} \ver^2
+ \frac{2 \lambda^2 m_B^6}{3 r} 
\vel \frac{A_2}{m_B+m_\rho} \ver^2 \nnb \\
&+& \frac{1}{2 r} m_B^2 \ga \frac{4}{3} \lambda  + 16 r s \dr 
\ga m_B + m_\rho \dr \vel A_1 \ver^2  -
\frac{4}{3} \lambda m_B^4 \frac{(1-r-s)}{r} A_1 A_2 \Bigg] \nnb \\
&+& \Bigg[ 2 \ga \mbox{\rm Im} \xi_1 \dr \eta_2 -
2 \ga \mbox{\rm Im} \lambda_u \dr \ga \mbox{\rm Im} \xi_2 \dr \eta_1 
+ 2 \ga \mbox{\rm Re} \lambda_u \dr \ga  \mbox{\rm Im} \xi_2 \dr \eta_2 \Bigg]
\nnb \\
&\times&\Bigg[ \frac{64 \lambda m_B^6 m_b s}{3 p^2} \, \frac{T_1 V}{m_B + m_\rho} +
\frac{8 \lambda^2 m_B^6 m_b}{3 r p^2}  \, \frac{A_2}{m_B + m_\rho}
\ga T_2 + \frac{p^2}{(m_B^2 - m_\rho^2)} \, T_3 \dr \nnb \\
&+& \frac{2 m_B^2 m_b}{r p^2} \ga \frac{4}{3} \lambda + 16 r s \dr 
A_1 T_2 ( m_B + m_\rho ) (m_B^2 - m_\rho^2) \nnb \\
&-& \frac{2}{3} \lambda m_B^4 (1-r-s) \Bigg( 
( m_B + m_\rho ) \frac{4 m_b}{p^2}\, 
A_1 \Big( T_2 + \frac{p^2}{(m_B^2 - m_\rho^2)} \, T_3 \Big) \nnb \\
&+& \frac{4 m_b ( m_B - m_\rho )}{p^2} \,A_2 T_2 \Bigg) \Bigg] \Bigg\}~, 
\eea
where 
\bea 
\Sigma^\rho &=& \frac{16}{3} \lambda m_B^6 s 
\ga \vel A \ver^2+ \vel C \ver^2 \dr + 
\frac{2}{3r} \lambda^2 m_B^6  \ga \vel B_2 \ver^2+ \vel D_2 \ver^2 \dr \nnb \\
&+& \frac{1}{2 r} \Bigg[ m_B^2 \ga \frac{4}{3} \lambda + 16 r s \dr 
\ga \vel B_1 \ver^2+ \vel D_1 \ver^2 \dr \nnb \\
&-& \frac{8}{3} \lambda m_B^4 (1-r-s) \Big( \ga \mbox{\rm Re} B_1 B_2 \dr
 + \ga \mbox{\rm Re} D_1 D_2 \dr \Big) \Bigg]~.
\eea
In deriving these expressions, we have used the following parametrization
\bea
C_9^{eff} &\equiv& \xi_1 + \lambda_u \xi_2~, \nnb \\
C_7^{eff} &\equiv& \eta_1 + i \,\eta_2~,
\eea
and assumed that all form factors are positive (see below). Interference of 
$C_9^{eff}$ and $C_7^{eff}$ terms gives new contribution to the CP
violating asymmetry. 
The results for the CP asymmetry in model II can be 
obtained from
Eqs. (26) and (27) by substituting Eq. (3) (i.e., $\eta_2 = 0$).  

At the end of this section we present forward--backward asymmetry $A_{FB}$,
which involve different combination of the Wilson coefficients. The analysis
of $A_{FB}$ is very useful in extracting precise information about the sign
of the Wilson
coefficients and the new physics. The forward--backward asymmetry is defined
as
\bea
A_{FB} (p^2) = \frac{\displaystyle{\int_0^1 dz \frac{d \Gamma}{dp^2 dz} -
\int_{-1}^0 dz
\frac{d \Gamma}{dp^2 dz}}}{\displaystyle{\int_0^1 dz \frac{d \Gamma}{dp^2
dz}+
\int_{-1}^0dz\frac{d \Gamma}{dp^2 dz}}}~.
\eea
The forward--backward asymmetry for the $B \rar \pi \ell^+ \ell^-$ decay is
zero, both in SM and 2HDM, in the limit $m_\ell \rar 0$. We can explain 
this fact briefly as follows. The hadronic current for $B \rar \pi \ell^+
\ell^-$ decay is a pure vector and the lepton current is also
conserved when $m_\ell \rar 0$. The charge asymmetry (or $A_{FB}$) is
non--zero if there exist $C$--vilolating terms but such terms are clearly 
absent in the $B \rar \pi \ell^+ \ell^-$. Using Eq. (24), the
forward--backward asymmetry for the $B \rar \rho\, \ell^+ \ell^-$ takes the
following form:
\bea
A_{FB}^\rho = \frac{8 m_B^4 s \sqrt{\lambda}
\Big[ \ga \mbox{\rm Re} B_1 C_1^* \dr + \ga \mbox{\rm Re} A_1 D_1^* \dr \Big]}
{\Sigma^\rho} ~.
\eea

Finally, we examine the CP--violating difference between $A_{FB}$ and 
$\bar A_{FB}$, i.e., 
\bea
\delta A_{FB} = A_{FB} - \bar A_{FB} \nnb~,
\eea
with $\bar A_{FB}$ being the forward--backward asymmetry in the 
anti particle channel, which can be obtained by the replacement
\bea
C_9^{eff}(\lambda_u) \rar \bar C_9^{eff}(\lambda_u \rar \lambda_u^*)~, \nnb
\eea
whose explicit can easily be obtained from Eq. (31), with the above
mentioned replacement of $C_9^{eff}$.

\section{Numerical analysis}
Before presentation of our quantitative calculations and graphics,
we would like to note that we have considered two different versions, namely
model II and model III of the 2HDM, in our analysis. For the free parameters
$\lambda_{bb}$ and $\lambda_{tt}$ of model III, we have used the
restrictions coming from $B \rar X_s \gamma$ decay, $B^0$--$\bar B^0$
mixing, $\rho$ parameter and neutron electric--dipole moment \cite{R28},
that yields $\vel \lambda_{bb} \ver = 50$, $\vel \lambda_{tt}\ver \le 0.03$.

The values of the main input parameteres, which appear in the expressions
for the branching ratios, $A_{FB}$ and $A_{CP}$ are: 
$m_b = 4.8~GeV,~m_c = 1.4~GeV,~m_\tau = 1.78~GeV,~m_B = 5.28~GeV,
m_\pi = 0.14~GeV$. For $B$ meson lifetime we take
$\tau(B) = 1.56 \time 10^{-12}~s$ \cite{R36}. 
The values of the Wilson coefficients are,
$C_1 = - 0.249$, $C_2 = 1.108$, $C_3 = 1.112 \times 10^{-2}$,
$C_4 = -2.569 \times 10^{-2}$, $C_5 = 7.4 \times 10^{-3}$,
$C_6 = - 3.144 \times 10^{-2}$. 
Throughout the course of the numerical analysis, we have used the
Wolfenstein parametrization of the CKM matrix elements, i.e., 
\bea
\lambda_u =\frac{V_{ub}V_{ud}^*}{V_{tb} V_{td}^*} =
\frac{\rho(1-\rho) -\eta^2 + i\, \eta}{(1-\rho)^2 + \eta^2} 
+ {\cal O}(\lambda^2)~, \nnb   
\eea
for which we have used the following three different sets of parameters,
\bea
(\rho,\eta) = \left\{ \begin{array}{l}
~~~(0.3;0.34) \\ \\
~~~(-0.07;0.34)\\ \\
~~~(-0.3;0.34)~.
\end{array} \right. \nnb
\eea
Of course the explicit expressions for the form factors are needed in the
present numerical analysis. In the current literature these form factors 
have been calculated in the
framework of the three point QCD sum rule \cite{R37}, relativistic quark
model \cite{R38}, and light  
cone QCD sum rules [39--41]. In further numerical analysis we have used
the light cone QCD sum rule predictions on the form factors. It should be
noted that the light cone QCD sum rule predictions on the form factors are
reliable in the region $m_b^2 - p^2 \sim {\cal O}$(few $GeV^2$). In order to
extend to the full physical region we have used best fitted expressions by
extrapolating the numerical results with the condition that these approximate   
formulas reproduce the light cone QCD sum rule predictions to a good
accuracy, in the above--mentioned region. The form of form factors which satisfy
this condition can be written in terms of three parameters as \cite{R39,R40}
\bea
F(p^2) = \frac{\ds {F(0)}}{\ds{1-a_F\,\frac{p^2}{m_B^2} + b_F \left
    ( \frac{p^2}{m_B^2} \right)^2}}~, \nnb
\eea
where the values of parameters $F(0)$, $a_F$ and $b_F$ for the relevant
decays, $B \rar \pi$ and $B \rar \rho$,
are listed in Table 1 (this Table is taken from \cite{R39,R40}

Firstly we consider model II for numerical calculations.
In Figs. (1) and (2) we present the dependence of the differential decay 
widths of the $B \rar \pi e^+ e^-$ and $B \rar \rho\, e^+ e^-$ on $p^2$ for 
$(\rho,\eta) = (0.3;0.34)$ at $m_{H^\pm} = 250~GeV$ and $tan\beta=1$, with and
without long distance contributions, correspondingly. 
In Figs. (3) and (4) we plot the variation of the CP--violating asymmetry 
$A_{CP}$ with respect to $p^2$, with the following set of parameters:
$(\rho,\eta) = (0.3;0.34)$ and $tan\beta=1$. In both figures, the solid
line corresponds to the SM case, dash--dotted and dotted lines represent 
the CP--violating asymmetry at two different values of the mass of 
charged Higgs boson $m_{H^\pm} = 250~GeV$ and $500~GeV$, respectively. The
total branching ratios for the $B \rar \pi e^+ e^-$ and $(B \rar \rho\, e^+ e^-)$
decays at three different sets of Wolfenstein parameters and at $m_{H^\pm} =
250 ~GeV$ are presented in Table 2.
From Figs. 1--4 we see that, in model II the dependences of the branching 
ratio and $A_{CP}$ 
asymmetry on $p^2$ are very similar to those predicted by the SM, but their
magnitudes different in these models. These results are expected, since in
model II, the charged Higgs contributions change only the values of the
Wilson coefficients $C_7$, $C_9$ and $C_{10}$. In this version of the 2HDM
charged Higgs contributions give rise to constructive interference to the SM
result. Therefore the branching ratio increases and CP asymmetry decreases. 

We presented in Table 2, the numerical values of the average values of the CP 
violating asymmetry $\la A_{CP} \ra$, in the
region $1~GeV^2 < p^2 < \ga m_{J/\psi} - 0.02~GeV \dr^2$, using the same values
of the Wolfenstein parameters used in Figs. 3 and 4.  
 
The dependence of the forward--backward asymmetry on $p^2$
$A_{FB}(B \rar \rho\, e^+ e^-)$ is plotted in Fig. (5) 
for SM and model II, with the same set of parameters as in Fig. (1).
It is observed that the value of $p^2$ at which $A_{FB}$ becomes zero is
shifted in model II. Therefore, in future experiments,
the determination of the value of $p^2$ at which 
$A_{FB}$ is zero can give unambiguous information about the presence of
new physics.
In Fig. (6) we plot the resulting difference in the forward--backward
asymmetry for the values of Wolfenstein parameters 
$(\rho,\eta) = (-0.07;0.2)$, with and without the long distance effects.
From these figures we observe that, $\delta F_{AB}$ for the 
$B \rar \rho\, e^+ e^-$ decay is positive in the non resonant region for all
values of $p^2$, both in SM and model II. 

Note that, the results we have presented for forward--backward
asymmetry and its difference are performed for $(\rho,\eta) = (-0.07;0.2)$.
However, for sake
of completeness, we have gone through the same analysis
for two different sets of the Wolfenstein parameters, namely, 
$(\rho,\eta) = (-0.3;0.34)$ and $(\rho,\eta) = (-0.07;0.34)$, as well as
several different choices of $\tan\beta$. The numerical
results and the relevant graphical presentations have demonstrated that, no 
remarkable differences have been observed among these different choices. 
In Figs. (7) and (8) we present the dependence of the CP asymmetry
$A_{CP}$, integrated over $p^2$, for the $B \rar \pi e^+ e^-$ and 
$B \rar \rho\, e^+ e^-$ decays on the phase angle $\theta$ at 
$m_{H^\pm} = 250~GeV$, $\vel \lambda_{bb}\ver = 50$ and 
$\vel \lambda_{tt}\ver = 0.03$,
without the long distance effects in model III. From both figures, especially from
$B \rar \rho\, e^+ e^-$ case,  we observe that, the average CP asymmetry 
differs essentially from the one predicted by model II. In the region
$\pi/2 < \theta < 3 \pi/2$, the change in $\left< A_{CP} \right>$ is more
than 2.5 times than that predicted by model II. This fact can be explained
as, the charged Higgs and SM contributions interference destructively in the
above--mentioned region of $\theta$. It should be stressed that, depending
on the value of the phase angle $\theta$, the charged Higgs contributions
can interfere with the SM results, either constructively or destructively.
This case is absolutely different in model II, where the above--mentioned
contributions interfere only constructively. The values of the branching
ratios $B \rar \pi \ell^+ \ell^-$ and $B \rar \rho \, \ell^+ \ell^-$ decays
at different values of the phase angle $\theta$ in model III are presented
in Table 4. 

In conclusion, the exclusive $B \rar \pi \ell^+ \ell^-$ and 
$B \rar \rho \, \ell^+ \ell^-$ decays are analyzed in the 2HDM and it is
found that, the CP violating asymmetry in model III differs essentially from
the ones predicted by model II.

\newpage

\section*{Figure captions}

{\bf Fig. 1} Invariant mass squared ($p^2$) distribution of the branching
ratio of the electron pair in the $B \rar \pi e^+ e^-$ decay. 
Line 1 corresponds to the mass spectrum including 
the effects of $\rho$, $\omega$ and $J/\Psi$ resonances, whereas line 2 
corresponds to the non resonant invariant mass spectrum, in the SM. Analogously, 
lines 3 and 4 represents the same distributions, respectively, in the model
II, at $tan \beta = 1$. In both models the Wolfenstein 
parameters are chosen to be $(\rho,\eta) = (0.3,0.34)$.\\ \\
{\bf Fig. 2} The same as in Fig. 1, but for the $B \rar \rho\, e^+ e^-$ decay.\\ \\
{\bf Fig. 3} CP--violating partial width asymmetry in the 
$B \rar \pi e^+ e^-$ decay as a function of $p^2$ for the values of the 
Wolfenstein parameters $(\rho,\eta) = (0.3,0.34)$, 
including $\rho$, $\omega$ and $J/\Psi$ resonances. Line 1 represents the SM. 
Lines 2 and 3 correspond to the model II case for the different choices of the 
charged Higgs 
boson mass $m_H^\pm = 500~GeV$, and $m_H^\pm = 250~GeV$, respectively.\\ \\
{\bf Fig. 4} The same as in Fig. 3, but for the $B \rar \rho\, e^+ e^-$ decay.\\ \\
{\bf Fig. 5} The dependence of the forward--backward asymmetry $A_{FB}$ on
$p^2$ in the $B \rar \rho\, e^+ e^-$ decay. The Wolfenstein parameters are 
chosen to be  
$(\rho,\eta) = (-0.07,0.2)$. See Fig. 1 for the interpretation
of the lines 1 to 4.\\ \\
{\bf Fig. 6} The CP--violating partial width asymmetry difference.
$\delta_{FB} = A_{FB} - \bar A_{FB}$ in the $B \rar \rho\, e^+ e^-$ decay for 
$(\rho,\eta) = (-0.07,0.2)$. See Fig. 1 for the interpretation of the lines 1 to 4. \\ \\   
{\bf Fig. 7} The dependence of the CP violating asymmetry, integrated over
$p^2$, on the phase angle $\theta$ for the $B \rar \pi e^+ e^-$
decay, in model III. In this figure the straight line corresponds to model II. 
The Wolfenstein
parameters and the charged Higgs are chosen to be $(\rho,\eta) = (0.3,0.34)$ and
$m_H^\pm = 250~GeV$, respectively.\\ \\
{\bf Fig. 8} The same as in Fig. 7, but for the $B \rar \rho\, e^+ e^-$ decay.
\newpage
\begin{figure}
\vskip 1.5cm
    \includegraphics{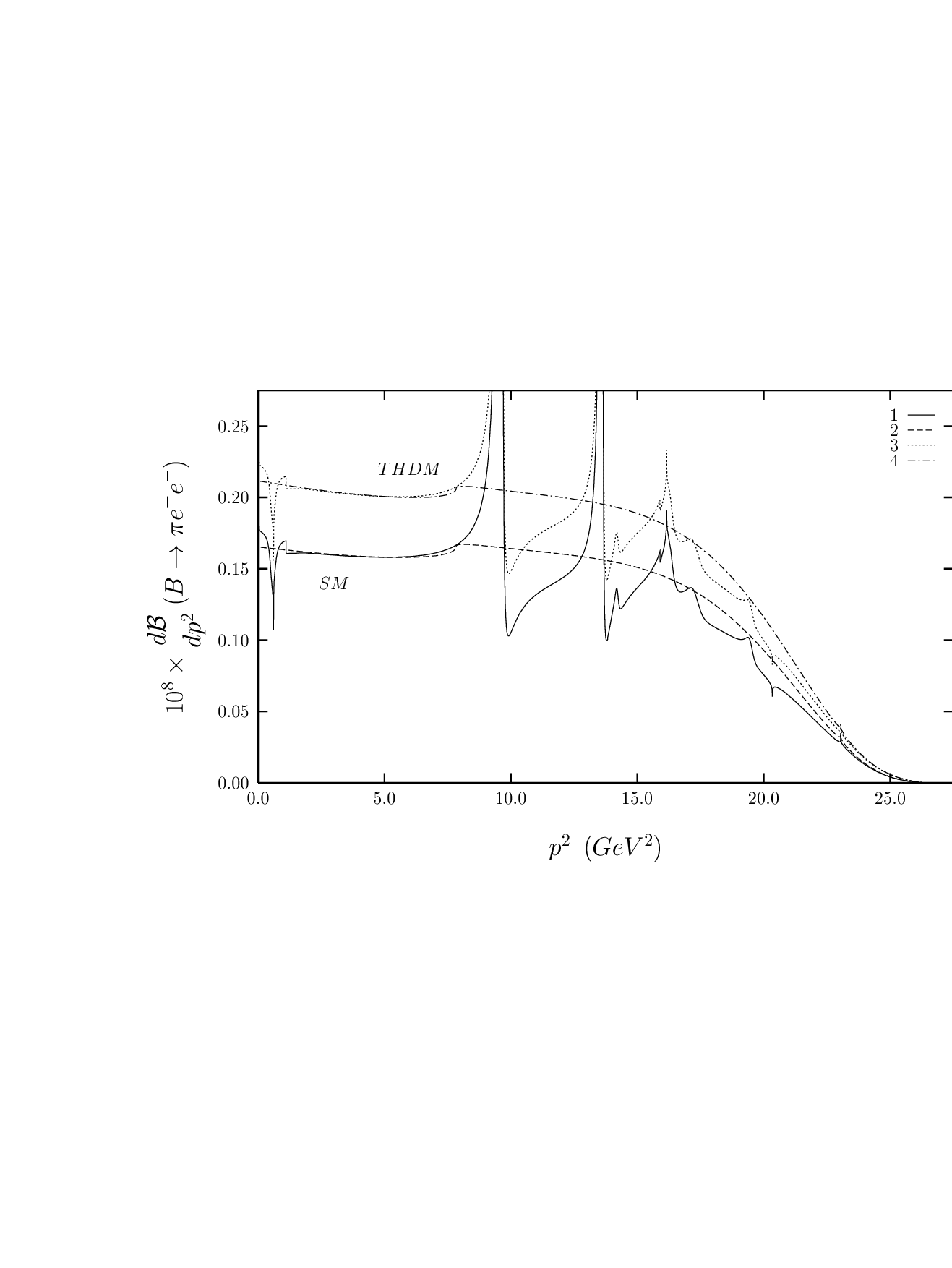}
\vskip 6.5cm  
\caption{}    
\end{figure}  
\begin{figure}
\vskip -1cm  
    \includegraphics{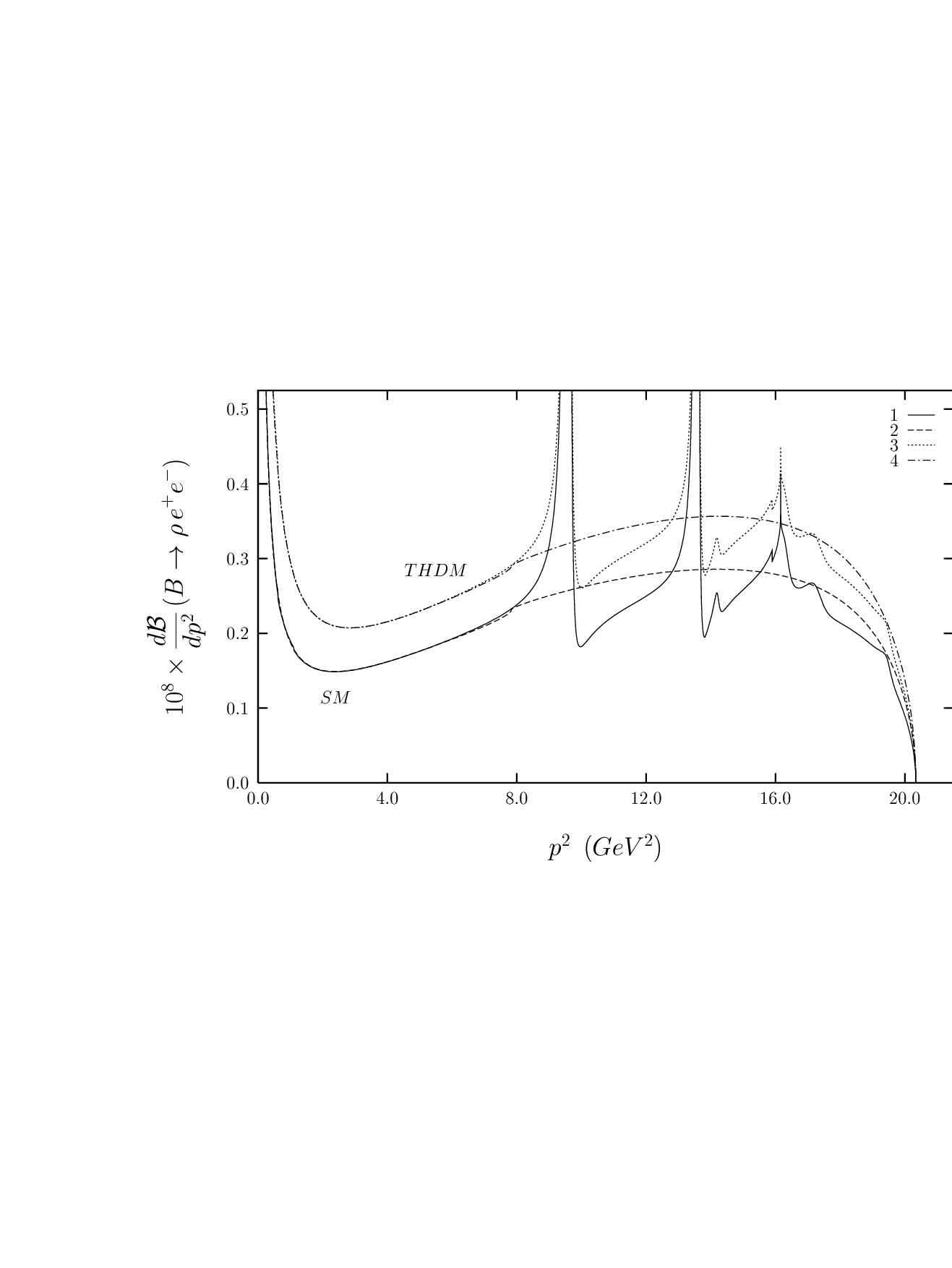}
\vskip 10cm
\caption{}
\end{figure}
\begin{figure}
\vskip 1.5cm
    \includegraphics{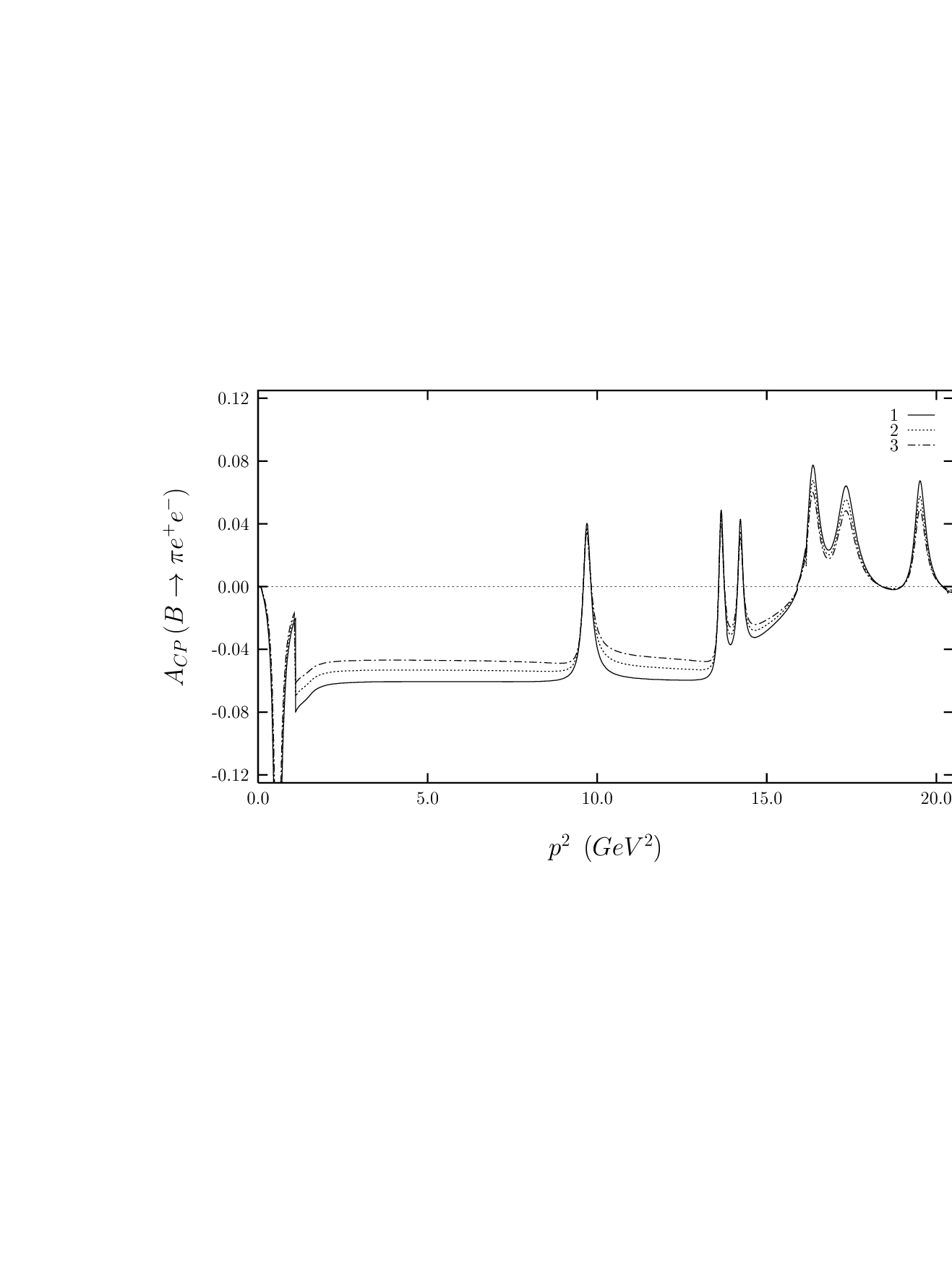}
\vskip 6.5cm  
\caption{}    
\end{figure}  
\begin{figure}
\vskip -1cm  
    \includegraphics{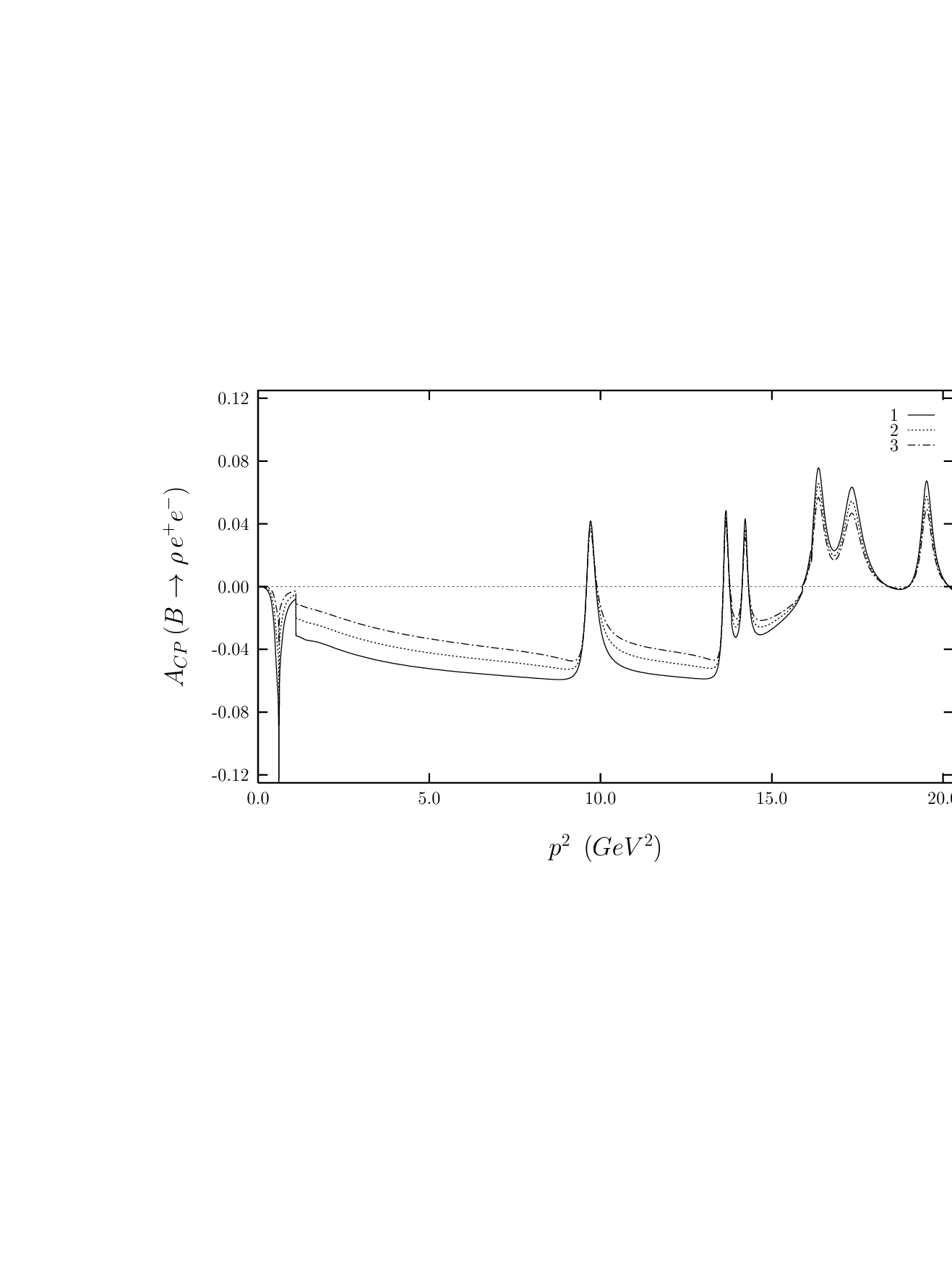}
\vskip 10cm
\caption{}
\end{figure}

\begin{figure}
\vskip 1.5cm
    \includegraphics{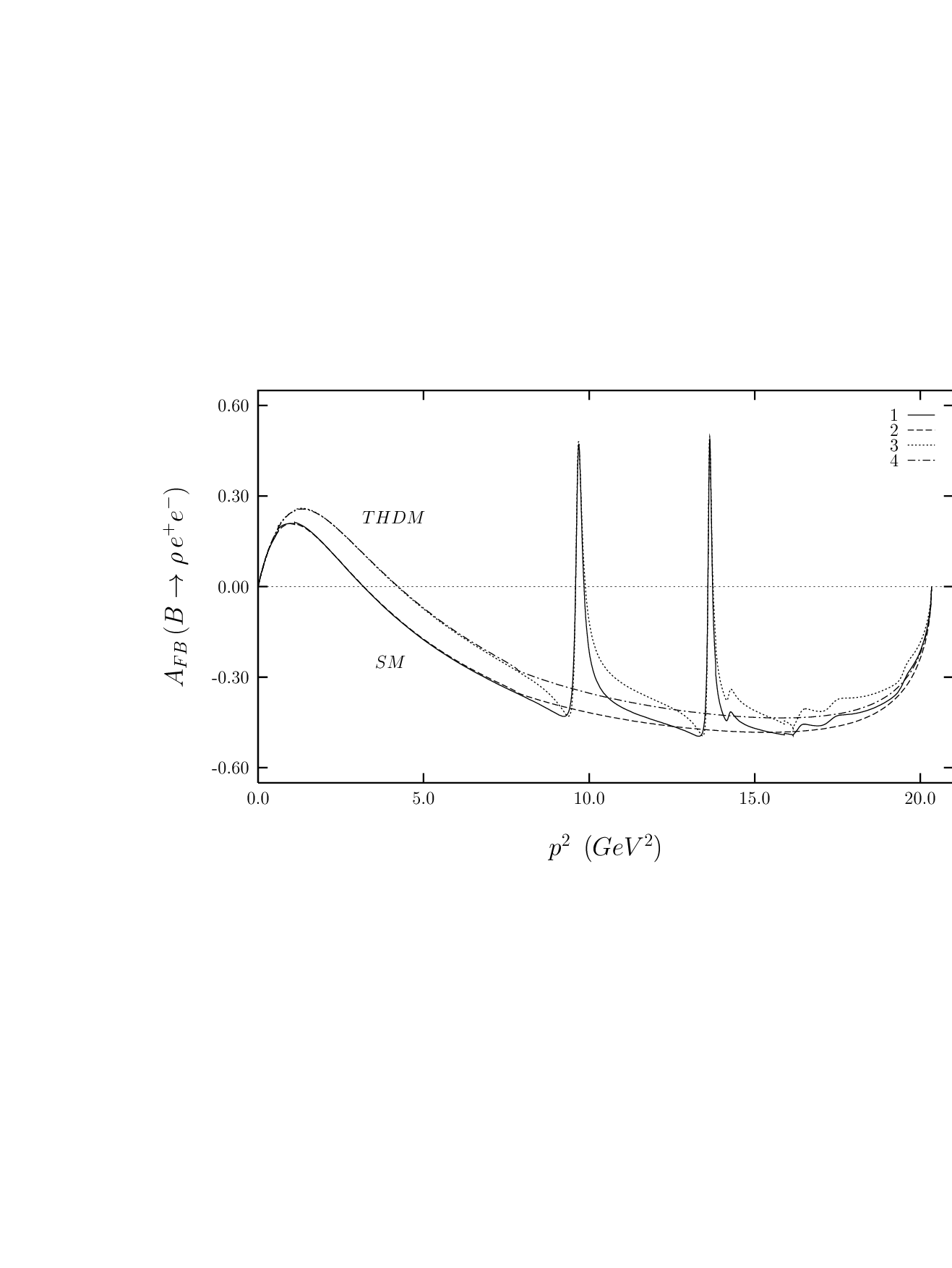}
\vskip 6.5cm  
\caption{}    
\end{figure}  
\begin{figure}
\vskip -1cm  
    \includegraphics{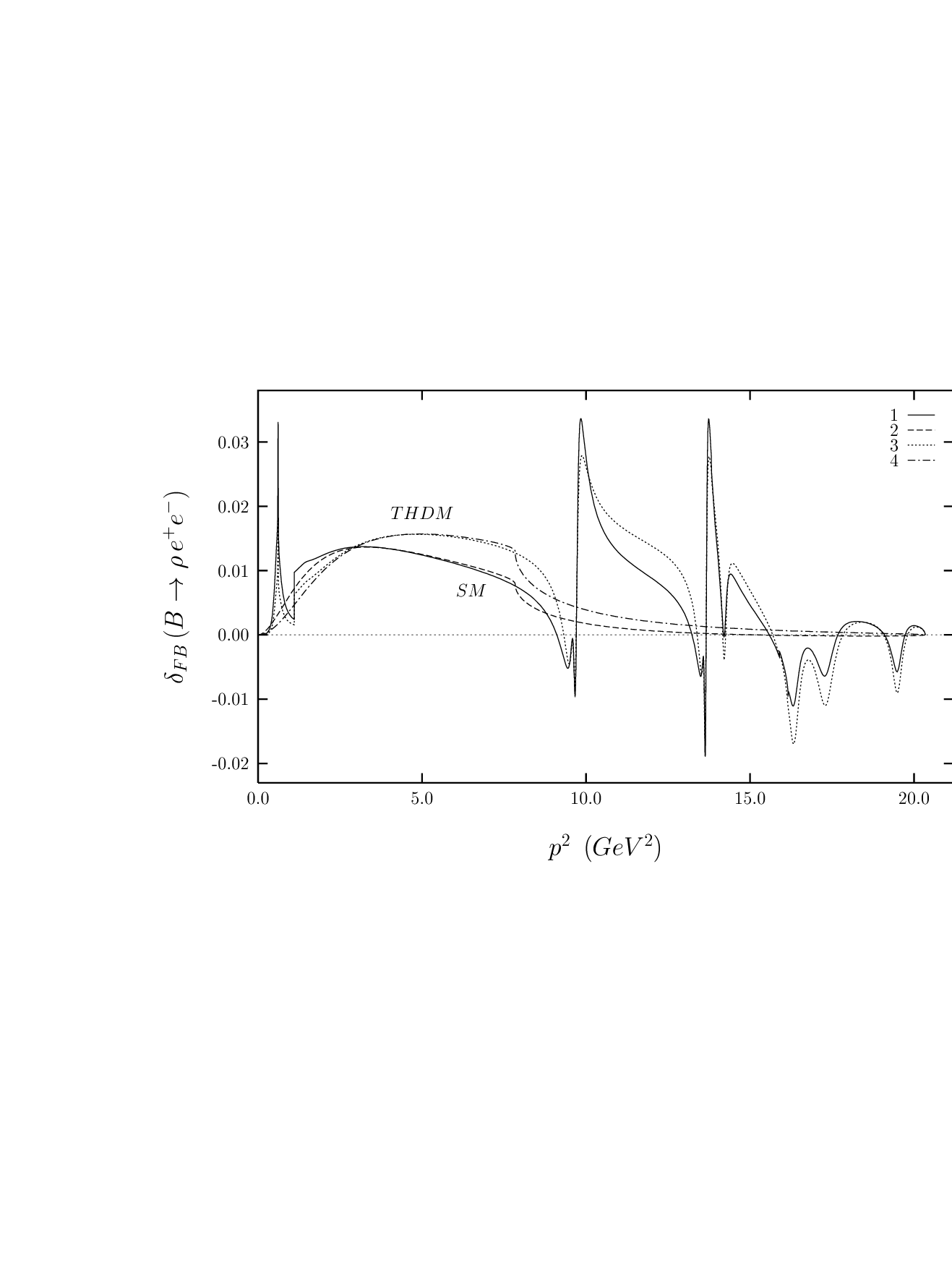}
\vskip 10cm
\caption{}
\end{figure}

\begin{figure}
\vskip 1.5cm
    \includegraphics{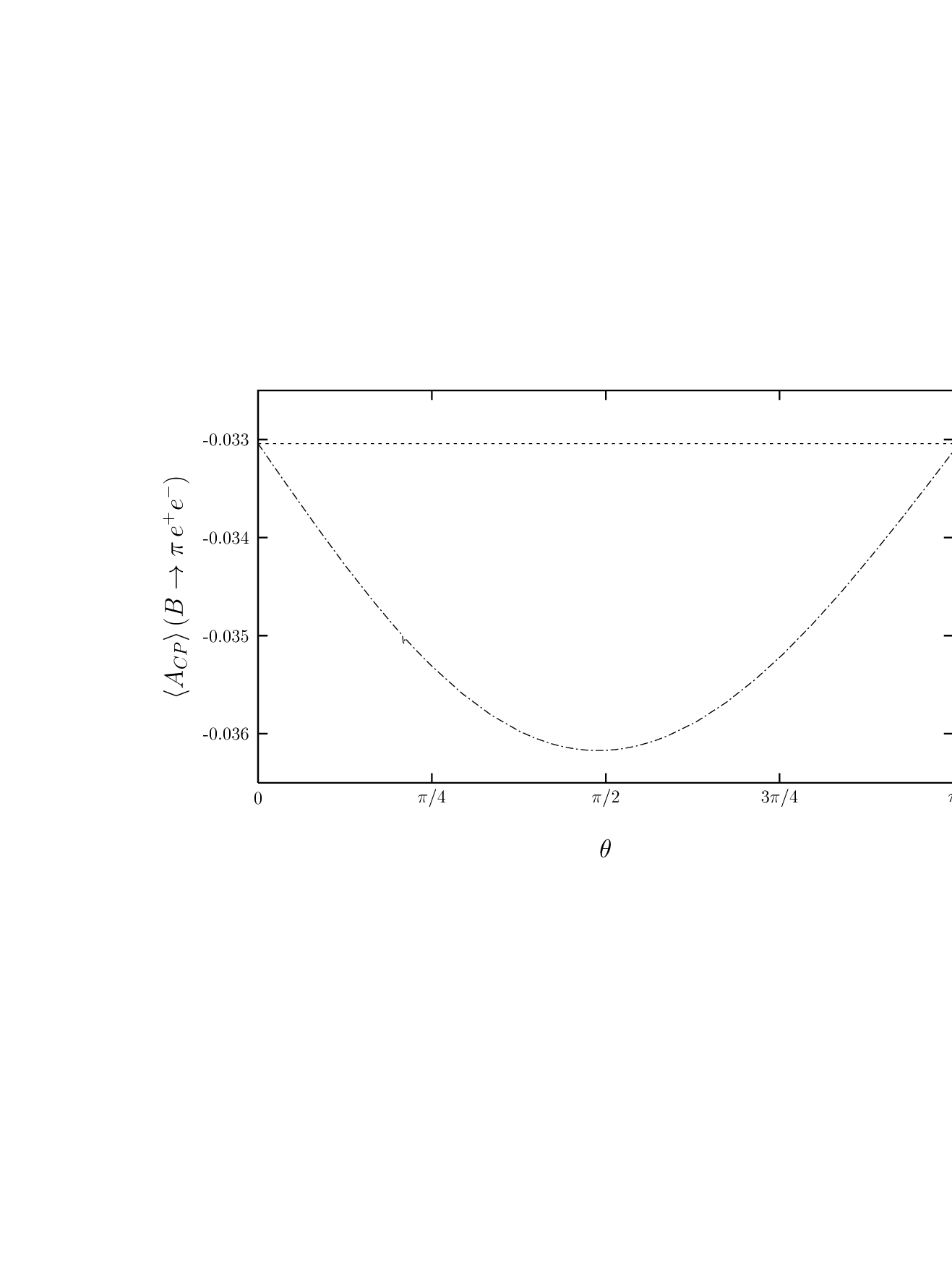}
\vskip 6.5cm  
\caption{}    
\end{figure}  
\begin{figure}
\vskip -1cm  
    \includegraphics{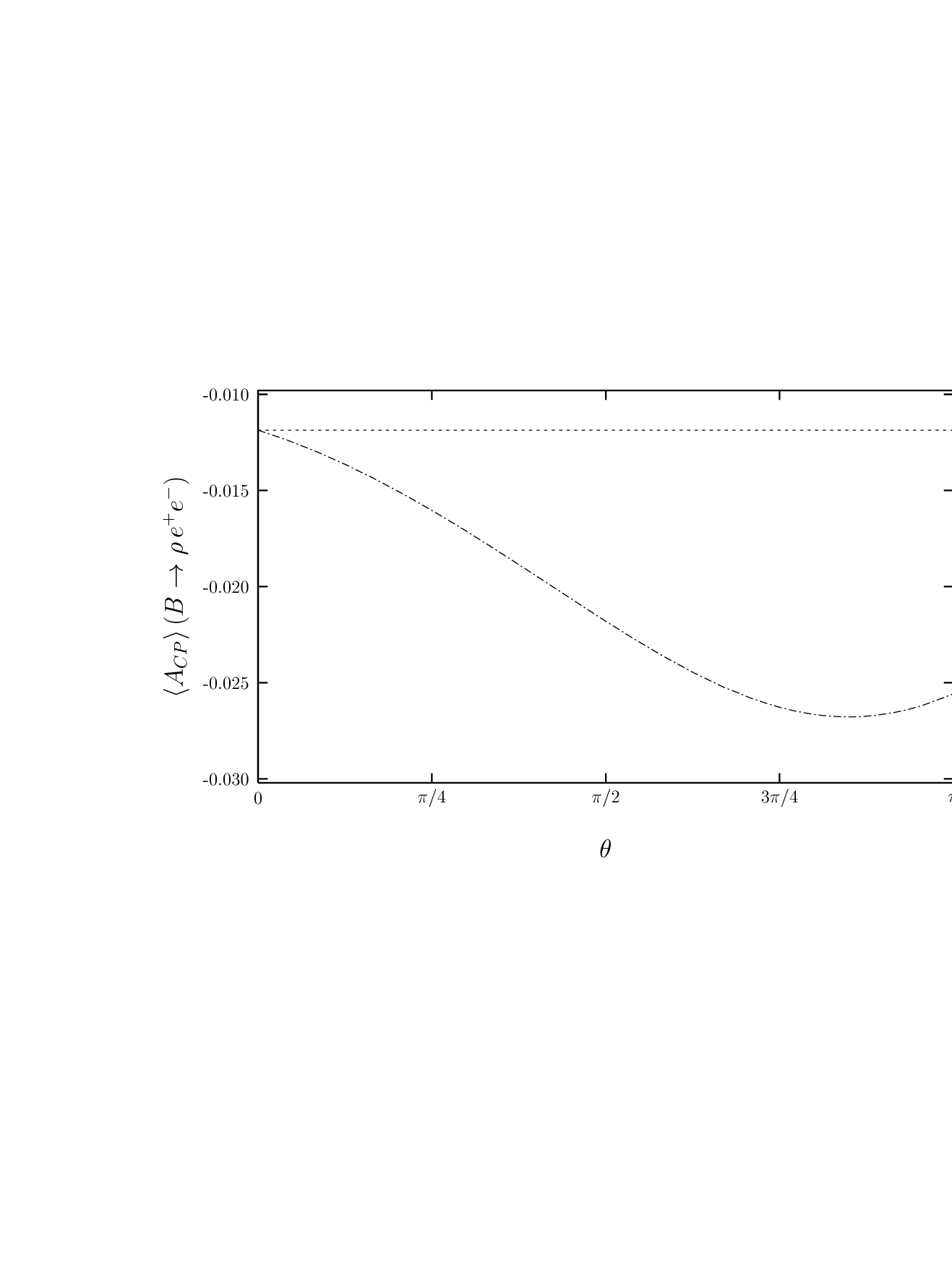}
\vskip 10cm
\caption{}
\end{figure}

\newpage
\begin{table}
\begin{center}
\begin{tabular}{|c|c|c|c|}
\hline
                   &\boss {$F(0)$} & $a_F$ & $b_F$  \\ \hline \hline
$~A_1^{B \rar\rho}~$ &\boss $\phantom{-}0.26 \pm 0.04$ & $0.29$
&$-0.415$\\ \hline
$~A_2^{B \rar\rho}~$ &\boss $\phantom{-}0.22 \pm 0.03$ & $0.93$
& $-0.092$\\ \hline
$~V^{B \rar\rho}~$ & \boss $\phantom{-}0.34 \pm 0.05$ & $1.37$
& $\phantom{-}0.315$\\ \hline
$~T_1^{B \rar\rho}~$ &\boss $\phantom{-}0.15 \pm 0.02$ & $1.41$                             
&$\phantom{-}0.361$ \\ \hline
$~T_2^{B \rar\rho}~$ &\boss $\phantom{-}0.15 \pm 0.02$ & $0.28$
&$-0.500$ \\ \hline
$~T_3^{B \rar\rho}~$ &\boss $\phantom{-}0.10\pm 0.02$ & $1.06$ 
&$-0.076$ \\ \hline \hline
$~f_+^{B \rar \pi }~$ &\boss $\phantom{-}0.30 \pm 0.04$ & $1.35$  
&$\phantom{-}0.270$ \\ \hline
$~f_T^{B \rar \pi }~$ &\boss $-0.30 \pm 0.04$ & $1.34$ & 
$\phantom{-}0.260$  
\\ \hline
\end{tabular}
\vskip 0.3 cm
\caption{}
\end{center}
\end{table}
\begin{table}[b]
\begin{center}
\begin{tabular}{|c|c|c|c|c|}
\hline
\multicolumn{1}{|c|}{\aaa }
&\multicolumn{2}{|c|}{${\cal B}(B \rar \pi e^+ e^-)$}
&\multicolumn{2}{|c|}{\dol{${\cal B}(B \rar \rho\, e^+ e^-)$}} \\
\hline\hline
$(\rho;~\eta)$ &\bos {SM} & THDM & SM & THDM \\  \hline
 (+0.3;~0.34)  &\aaa $3.27 \times 10^{-8}$ & $4.11 \times 10^{-8}$
& $5.99 \times 10^{-8}$ &$\dol{8.45 \times 10^{-8}}$ \\ \hline
(-0.3;~0.34)  &\aaa $3.31 \times 10^{-8}$ & $4.15 \times 10^{-8}$
& $ 6.00\times 10^{-8}$ & $\dol{8.46 \times 10^{-8}}$ \\ \hline
{(-0.07;~0.34)}  & \aaa $3.30 \times 10^{-8}$ & $4.14 \times 10^{-8}$
& $6.00 \times 10^{-8}$ & $\dol{8.46 \times 10^{-8}}$ \\ \hline
\end{tabular}
\vskip 0.3 cm
\caption{}
\end{center}
\end{table}
\newpage
\begin{table}
\begin{center}
\begin{tabular}{|c|c|c|}
\hline
&\boss {$\left< A_{CP} \right>^{(B \rar \pi)}$} & 
$\left< A_{CP} \right>^{(B \rar \rho)}$   \\ \hline \hline
$m_{H^\pm} = 250~GeV$ &\boss {-0.048} & -0.030 \\ \hline
$m_{H^\pm} = 500~GeV$  &\boss {-0.054} & -0.031 \\ \hline
\end{tabular}
\vskip 0.3 cm
\caption{}
\end{center}
\end{table}
\begin{table}
\begin{center}
\begin{tabular}{|c|c|c|}
\hline
$\theta$&\boss {${\cal B}(B \rar \pi e^+ e^-)$} & 
${\cal B}(B \rar \rho\, e^+ e^-)$   \\ \hline \hline
$0$              &\boss {$3.12 \times 10^{-8}$} & $7.41 \times 10^{-8}$ \\ \hline
$\pi/4$  &\boss {$3.16 \times 10^{-8}$} & $7.08 \times 10^{-8}$ \\ \hline
$\pi/2$  &\boss {$3.26 \times 10^{-8}$} & $6.34 \times 10^{-8}$ \\ \hline
\end{tabular}
\vskip 0.3 cm
\caption{}
\end{center}
\end{table}

\end{document}